\documentclass{sf2a-conf2022}
\usepackage{graphicx}
\usepackage{epstopdf}
\usepackage{natbib}
\usepackage[citecolor=blue, linkcolor=blue, urlcolor = black, colorlinks = true, backref = page]{hyperref}
\usepackage{amsmath}
\usepackage{graphicx}

\def\BibTeX{{\rm B\kern-.05em{\sc i\kern-.025em b}\kern-.08em
    T\kern-.1667em\lower.7ex\hbox{E}\kern-.125emX}}
\bibpunct{(}{)}{;}{a}{}{,}  

\begin{document}

\TitreGlobal{SF2A 2022}


\title{Can we detect deep axisymmetric toroidal magnetic fields \\in stars?}

\runningtitle{Can we detect deep axisymmetric toroidal magnetic fields in stars?}

\author{H. Dhouib}\address{Université Paris Cité, Université Paris-Saclay, CEA, CNRS, AIM, F-91191, Gif-sur-Yvette, France}
\author{S. Mathis$^{1}$}
\author{L. Bugnet}\address{Flatiron Institute, Simons Foundation, 162 Fifth Ave, New York, NY 10010, USA}
\author{T. Van Reeth}\address{Institute of Astronomy, KU Leuven, Celestijnenlaan 200D, 3001 Leuven, Belgium}
\author{C. Aerts$^{3,4,}$}
\address{Department of Astrophysics, IMAPP, Radboud University Nijmegen, PO Box 9010, 6500 GL Nijmegen, the Netherlands} 
\address{Max Planck Institute for Astronomy, Koenigstuhl 17, 69117 Heidelberg, Germany}

\setcounter{page}{237}


\maketitle

\begin{abstract}
One of the major discoveries of asteroseismology is the signature of a strong extraction of angular momentum (AM) in the radiative zones of stars across the entire Hertzsprung-Russell diagram, resulting in weak core-to-surface rotation contrasts. Despite all efforts, a consistent AM transport theory, which reproduces both the internal rotation and mixing probed thanks to the seismology of stars, remains one of the major open problems in modern stellar astrophysics.
A possible key ingredient to figure out this puzzle is magnetic field with its various possible topologies. Among them, strong axisymmetric toroidal fields, which are subject to the so-called Tayler MHD instability, could play a major role. They could trigger a dynamo action in radiative layers while the resulting magnetic torque allows an efficient transport of AM.
But is it possible to detect signatures of these deep toroidal magnetic fields?
The only way to answer this question is asteroseismology and the best laboratories of study are intermediate-mass and massive stars because of their external radiative envelope. Since most of these are rapid rotators during their main-sequence, we have to study stellar pulsations propagating in stably stratified, rotating, and potentially strongly magnetised radiative zones.
For that, we generalise the traditional approximation of rotation, which provides in its classic version a flexible treatment of the adiabatic propagation of gravito-inertial modes, by taking simultaneously general axisymmetric differential rotation and toroidal magnetic fields into account.
Using this new non-perturbative formalism, we derive the asymptotic properties of magneto-gravito-inertial modes and we explore the different possible field configurations. We found that the magnetic effects should be detectable for equatorial fields using high-precision asteroseismic data.
\end{abstract}

\section{Introduction}
Asteroseismology has revealed a strong signature of extraction of angular momentum (AM) operating in the radiative zones of stars of all types throughout their evolution, leading to weak core-to-surface rotation contrasts \citep[e.g.][]{Beck2012RG,Mosser2012,Deheuvels2012,Deheuvels2014,Aerts2017,vanreeth2016,vanreeth2018,Li2020,Saio2021}.
The current angular momentum transport models fail to reproduce these astroseismic observations \citep[e.g.][]{Ceillier2013,Marques2013,Cantiello2014,ouazzani2019}.
One of the plausible candidates to efficiently carry AM and explain this discrepancy is magnetic fields in radiative zones of early-type stars. One of the interesting magnetic field configurations is initially axisymmetric toroidal fields which was studied by \cite{spruit2002}. He argued that such field could trigger a dynamo action in stably stratified layers which can cause a magnetic torque
that allows a very efficient transport of AM. 
The way the dynamo loop in stellar radiative zones is closed  has been strongly debated in the literature \cite[e.g.][]{Zahnetal2007, Fulleretal2019}. A first proof of how such dynamo can be driven by a combination of the action of differential rotation and the Tayler instability \citep{Tayler1973} was found using global 3D numerical simulations by \cite{Petitdemange2022}.
However if a strong axisymmetric toroidal magntic field is present in the radiative envelope but does not emerge at the stellar surface as found in this latter work, spectropolarimetry is blind. So, how they can be detected and characterised?
The only diagnostic tool that we have to access this information is magneto-asteroseismology \citep[e.g.][]{Neiner2015,Mathisetal2021, Bugnet2021,Lecoanet2022,Bugnet2022} which consists of searching for magnetic signatures in the stellar oscillations frequency spectrum.
This was first done using a perturbative approach by \cite{Prat2019, Prat2020} and \cite{VanBeeck2020} who studied the impact of axisymmetric and inclined mixed (i.e.\ poloidal plus toroidal) dipolar fossil fields on the frequency spectrum of gravito-inertial modes. These oscillations propagate in rotating stellar radiative zones under the combined action of the buoyancy force and the Coriolis acceleration). They found a magnetic signatures (sawtooth patterns) that differ appreciably from those due to rotation and chemical stratification \citep{Miglio2008,vanreeth2018}.
We study here the impact of general axisymmetric toroidal magnetic fields on gravito-inertial modes in a non-perturbative way by generalising the traditional  approximation of rotation (TAR). This approximation was first introduced in geophysics by \cite{eckart1960} and intensively used in asteroseismology for the study of low-frequency waves propagating in rotating strongly stratified zones \citep[e.g.][]{Bouabidetal2013}. It is built on the assumptions that the modes have a low frequency and propagate in strongly stratified regions which allows us to neglect of the vertical component of the Coriolis acceleration. Therefore, the vertical and the horizontal dynamics become separable. But the standard version of this approximation supposes that the star is spherical, uniformly rotating and neglect the magnetic field. We derive here a new generalisation of this formalism that allow us to take into account general toroidal magnetic fields in a non-perturbative way.
Next, we derive seismic diagnosis used intensively in gravito-inertial modes asteroseimology \citep[e.g.][]{aerts2021}. Then, we discuss the detectability of such fields by the space missions Kepler, K2, TESS, and in the future PLATO.

\section{Magnetic TAR}
To study the effect of the  magnetic fields on gravito-inertial modes, we derive the magnetic TAR, which includes a general axisymmetric toroidal magnetic field (with a general 2D profile of the Alfvén frequency $\omega_{\rm A}$) in a non-perturbative way and a radial differential rotation $\Omega$.  Using this formalism we derive the magnetic Laplace tidal equation \citep{laplace1799}
    \begin{equation}
    \mathcal{L}^{\rm{magn.}}_{\nu m} \left[w^{\rm{magn.}}_{\nu k m} \right] = -\Lambda^{\rm{magn.}}_{\nu k m}(r) w^{\rm{magn.}}_{\nu k m}(r,x),
    \end{equation}
such that
\begin{equation}
    \mathcal{L}^{\rm{magn.}}_{\nu m} = \omega^2 \partial_x \left[\frac{1}{\mathcal{A}}\frac{1-x^2}{D_{\rm M}}\partial_x\right] +
    m\omega^2 \partial_x\left(\frac{ \nu_{\rm M} x}{\mathcal{A}D_{\rm M}}\right) -m^2\frac{\omega^2}{\mathcal{A}D_{\rm M}\left(1-x^2\right)} + m^2 \frac{\omega^2}{\mathcal{A}^2}\frac{x}{D_{\rm M}} \partial_x \omega_{\rm A}^2,
\end{equation}
where $x=\cos{\theta}$ the reduced latitudinal coordinate, $m$  the azimuthal order, $k$ the index of a latitudinal eigenmode, $\omega$ the wave frequency in the rotating frame,
$\nu_{\rm M} = \mathcal{B}/\mathcal{A}$ the magnetic spin parameter, $\nu  = 2\Omega/\omega$ the standard rotation spin parameter, $\mathcal{A} =\omega^{2}-m^{2} \omega_{\rm A}^{2}$, $\mathcal{B} =2\left(\Omega \omega+m \omega_{\rm A}^{2}\right)$, and $D_{\rm M}  = 1- \nu_{\rm M}^2 x^2-\left(1-x^2\right)\frac{x}{\mathcal{A}}\partial_x\omega_{\rm A}^2$. $w_{\nu k m}^{\rm{magn.}}$ and $\Lambda_{\nu k m}^{\rm{magn.}}$ represent, respectively, the magnetic generalised Hough functions and the eigenvalues.
The standard Laplace tidal equation is altered and gets a radial dependence, propagating into solutions that are modified relative to those found in the hydrodynamic case \citep{Bildsten1996,lee+saio1997}.

Using the radial quantification relation \citep{unno1989}, we can derive the analytical expression of the asymptotic frequencies of low-frequency magneto-gravito-inertial modes in the rotating frame
\begin{equation}\label{eq:frequencies}
    \omega_{n k m}=\frac{1}{(n+1 / 2) \pi} \int_{r_{1}}^{r_{2}} \frac{N\left(r\right)}{r}\sqrt{\Lambda_{\nu_{n} k m}^{\rm{magn.}}\left(r\right)}\;\mathrm{d} r,
\end{equation}
where $n$ is the radial order and $r_1$ and $r_2$ are the turning points of the 
Brunt–Väisälä frequency $N$. We obtain a similar expression to the one found in the non-magnetic case. The difference is that the eigenvalues are modified due to the magnetic field (no longer constant but depend on the radius). 
This equation is implicit because $\Lambda_{\nu_{n} k m}^{\rm{magn.}}$ depends on $\nu_{n}$ which in turn depends on the frequency $\omega_{n k m}$. Therefore, Eq.\;\eqref{eq:frequencies} must be solved numerically.

\section{Detectability of equatorial and hemispheric toroidal magnetic fields}
\begin{figure}[ht!]
 \centering
\includegraphics[width=0.48\textwidth,clip]{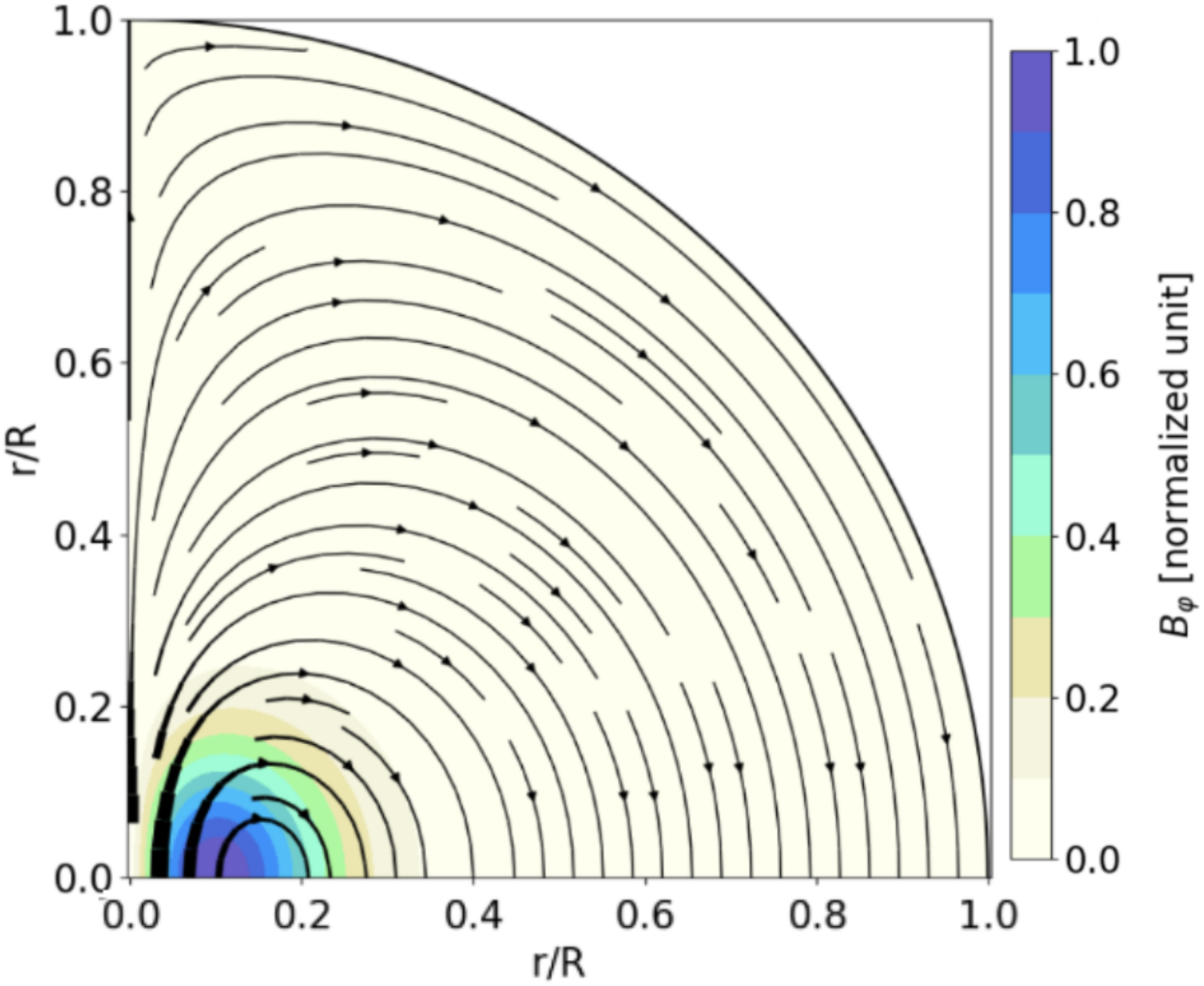}      
\includegraphics[scale=0.20,clip]{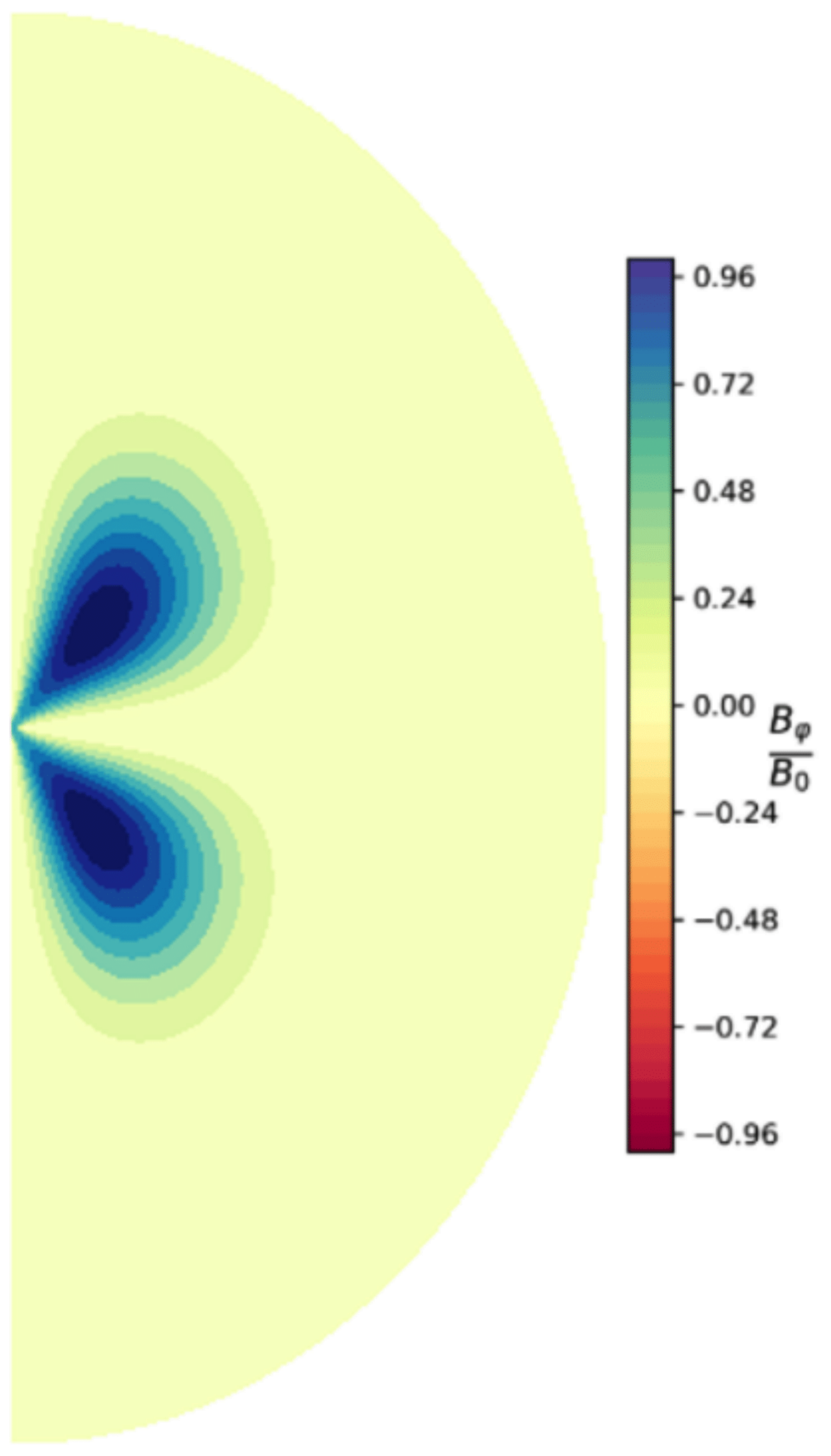}  
  \caption{Equatorial {\bf (Left)} and hemispheric {\bf (Right)} toroidal magnetic configurations.}
  \label{fig:field}
\end{figure}

\begin{figure}[ht!]
 \centering
\includegraphics[width=0.48\textwidth,clip]{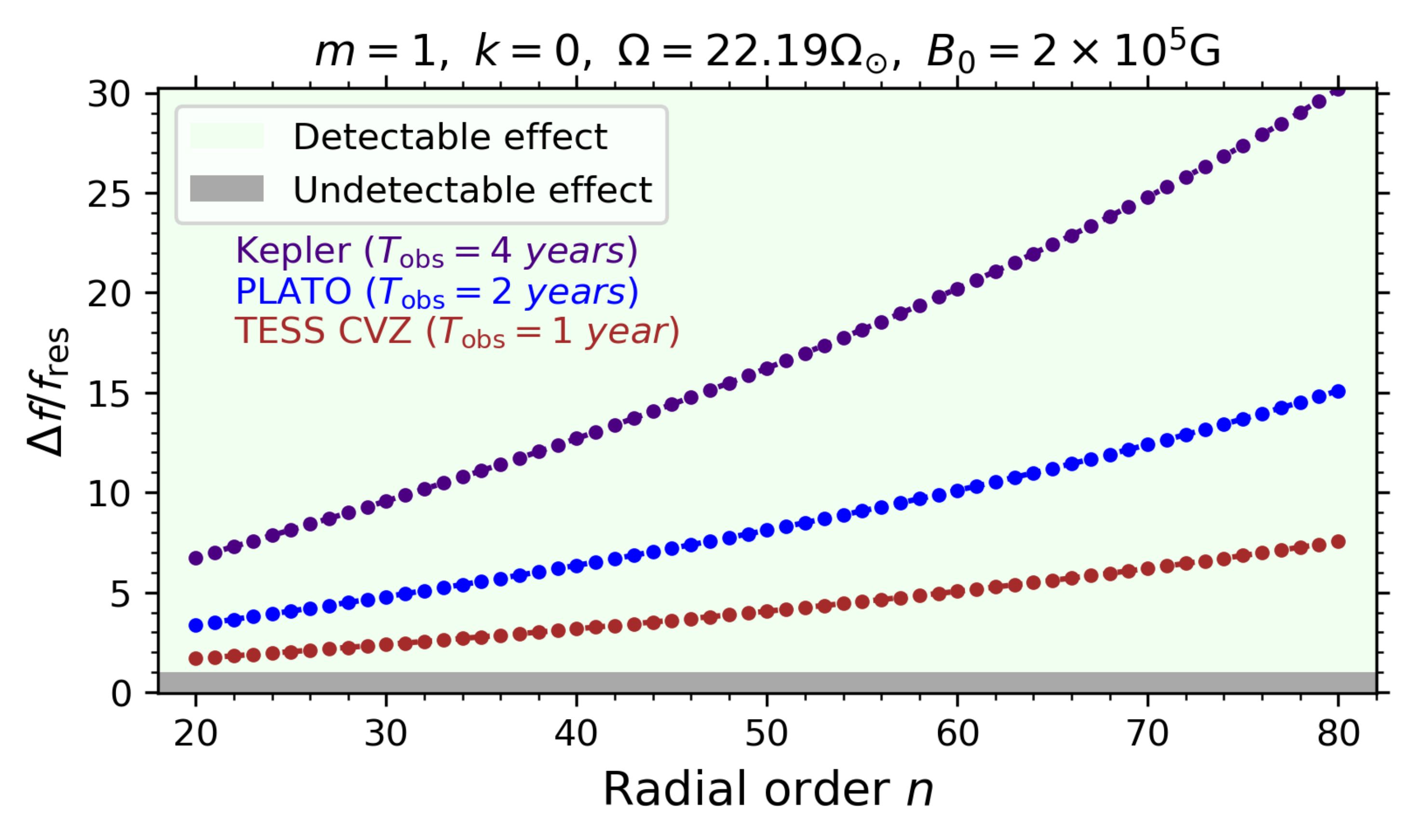}      
\includegraphics[width=0.48\textwidth,clip]{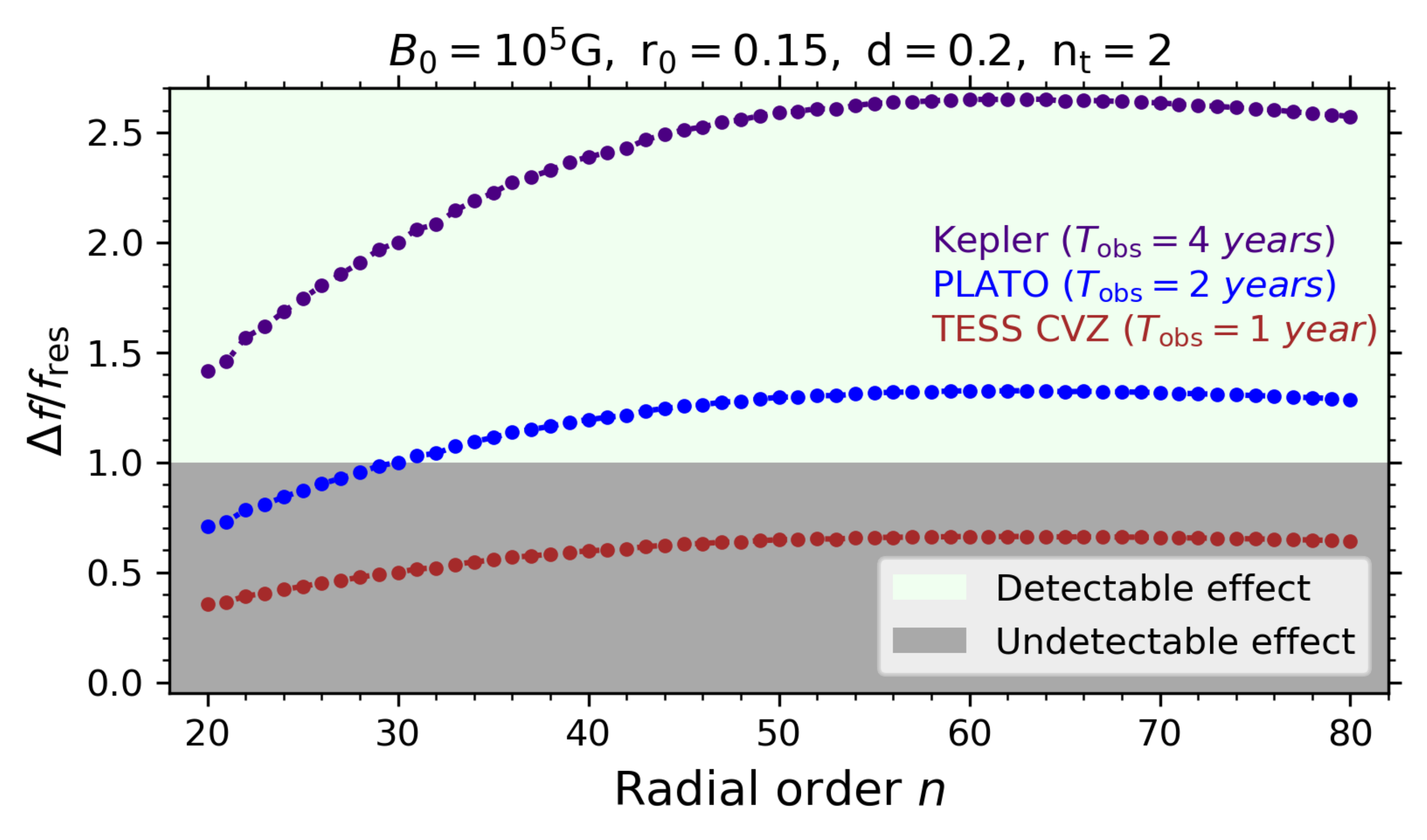}  
  \caption{Detectability of the effect of an equatorial {\bf (Left)} and hemispheric {\bf (Right)} toroidal magnetic fields on the $\{k=0,m=1\}$ mode as a function of the radial order  $n$ using a $1.6{\rm M}_\odot$ $\gamma\,$Dor model near ZAMS.}
  \label{fig:detect}
\end{figure}

As a proof of concept, we applied the magnetic TAR to representative stellar models of typical $\gamma$\,Dor stars during their main-sequence evolution (see \cite{dhouib2022} for the complete exploration).
We investigate the impact of two different configurations of toroidal magnetic field. The first one is an equatorial field localised near the equator, similar to the one found in the numerical simulations of \cite{Petitdemange2022}, represented in Fig.\;\ref{fig:field} (left) for a $\gamma$\,Dor star at ZAMS. We can see that the magnetic field is the strongest in the inner region of the radiative zone, whereas near the surface it becomes very weak.
The second is a hemispheric field represented in Fig.\;\ref{fig:field} (right) composed of two magnetic distributions localised in the Northern and Southern hemispheres and vanishing at the equator \citep{Jouve2015}. 
As we can see in Fig.\;\ref{fig:detect}, the effect of an equatorial magnetic field is in principle largely detectable for all radial orders using nominal TESS CVZ, \textit{Kepler} and PLATO light curves. But the detection becomes much harder for hemispheric fields since most gravito-inertial modes are trapped in the equatorial belt. Therefore, the modes are weakly impacted by this magnetic field because it is localised far from the equator.
Despite the fact that a toroidal magnetic field is theoretically detectable, it lacks a distinct signature in the period spacing pattern. It instead introduces mode period shifts that occur concurrently with those caused by other physical effects such as centrifugal acceleration or differential rotation \citep{dhouib2021a, dhouib2021b}.
The degeneracy among astrophysical phenomena can mask the magnetic effect in forward asteroseismic modelling analyses. In particular in the case of purely axisymmetric toroidal fields for which the Lorentz force has the same mathematical structure as the Coriolis acceleration. In order to tackle this problem, work is being  \citep{vanreeth2018, Mombarg2022} and will be done to improve the modelling of these processes in order to unravel and distinguish the various signatures.

\section{Conclusions}
Even though the toroidal configurations studied here are unstable \citep{Tayler1973, Markey1973}, they are of high interest since they can be sustained by a dynamo action generated by the differential rotation and the  Tayler instability which can efficiently transport AM.
Here we test the detectability of these fields by asteroseismology. We found that the effect of an equatorial magnetic field is in theory largely detectable for all radial orders. In contrast, the detection is much difficult for hemispheric fields because most gravito-inertial modes are trapped in the equatorial belt.
This work is the first step towards a general MHD TAR with a complex mixed poloidal and toroidal field which corresponds to stable fossil field configurations \citep[e.g.][]{Braithwaite2004, Duez2010}.

\bibliographystyle{aa}  
\bibliography{Dhouib} 
\end{document}